\documentclass[twocolumn,prb,aps,superscriptaddress]{revtex4}
\usepackage{graphicx}
\usepackage{enumitem}
\usepackage{bm}

\newcommand{\TaS}{$2H$-TaS$_2$\,}
\newcommand{\NbSe}{$2H$-NbSe$_2$\,}
\newcommand{\TaSe}{$2H$-TaSe$_2$\,}

\newcommand{\Tcdw}{$\it{T}_{\text{cdw}}\,$}
\newcommand{\qcdw}{$\textbf{q}_{\text{cdw}}\,$}

\newcommand{\wbar}{$\overline{\omega}$\,}

\begin{document}

\title{Spectroscopic signature of the $\mathbf{k}$-dependent electron-phonon coupling in \TaS}
\author{K. Wijayaratne}
\affiliation{Department of Physics, University of Virginia, Charlottesville, VA 22904, USA}
\author{J. Zhao}
\affiliation{Department of Physics, University of Virginia, Charlottesville, VA 22904, USA}
\author{C. D. Malliakas}
\affiliation{Department of Chemistry, Northwestern University, Evanston, IL 60208, USA}
\author{D. Y. Chung}
\affiliation{Materials Science Division, Argonne National Laboratory, Argonne, IL 60439, USA}
\author{M. G. Kanatzidis}
\affiliation{Department of Chemistry, Northwestern University, Evanston, IL 60208, USA}
\affiliation{Materials Science Division, Argonne National Laboratory, Argonne, IL 60439, USA}
\author{U. Chatterjee}\thanks{Correspondence to: Email: uc5j@virginia.edu (U.C.)}
\affiliation{Department of Physics, University of Virginia, Charlottesville, VA 22904, USA}

\begin{abstract}

Our detailed Angle Resolved Photoemission Spectroscopy (ARPES) study of \TaS, a canonical incommensurate charge density wave (CDW) material, illustrates pronounced many-body renormalization in the system, which is manifested by the presence of multiple kink structures in the electronic dispersions. Temperature-dependent measurements reveal that these kink structures persist even at temperatures higher than the charge density wave transition temperature \Tcdw and the energy locations of the kinks are practically  temperature-independent. Correlating kink energies with the published Raman scattering data and the theoretically calculated phonon spectrum of \TaS, we conclude phononic mechanism for these kinks. We have also detected momentum-anisotropy in the band renormalization, which in turn indicates momentum-dependence of the electron-phonon coupling of the system.
\end{abstract} 

\maketitle

\section{Introduction}
Many-body effects and their subtle interplay often lead to fascinating physical phenomena in solid state systems. Examples include superconductivity in cuprate high temperature superconductors (HTSCs) \cite{HTSCs_REF_1, HTSCs_REF_2, HTSCs_REF_3}, unusual mass renormalization in heavy fermion compounds \cite{HEAVY_FERMION_REF_1, HEAVY_FERMION_REF_2}, colossal magnetoresistance (CMR) in manganites \cite{CMR_REF_1, CMR_REF_2} and charge density wave (CDW) in various $2H$ and $1T$ transition metal dichalcogenides (TMDs) \cite{TMD_REF_1, TMD_REF_2, TMD_REF_3}. A prototypical example of a many-body interaction is the electron-phonon (el-ph) interaction, whose significance can't be overemphasized  in TMDs, particularly in  $2H$ polytypes of TMDS such as  \TaSe, \NbSe and \TaS (hereinafter collectively referred to as $2H$ TMDs). The reason is that the el-ph coupling is central to both superconductivity \cite{BCS_PAPER} and CDW \cite{JVW_NATCOM, JVW_PRB, VARMA_STRONG_COUPLING}, which either coexist or compete in these materials.

Despite decades of intense research, the mechanism behind CDW order in $2H$ TMDs remains an intriguing puzzle. Traditionally, their CDW orders have been conjectured to be analogous to the Peierls instability occurring due to the so-called Fermi Surface (FS) nesting \cite{PEIERLS_INSTABILITY}. The FS's of \TaSe, \NbSe and \TaS indeed have extended nearly parallel regions, which in turn could lead to  CDW orders mediated by FS nesting. However, the CDW wave-vectors (\qcdw's) of these compounds have been shown to be incompatible with their FS nesting vectors, which are the vectors spanning the mutually parallel sections  of their FS's \cite{BORISENKO_NbSe2_PRL, UC_NATCOM, KYLE_NbSe2_PRB_OLD, UC_TaS2}. An alternative theoretical idea \cite{RICE_SCOTT_SADDLE_BAND} relate the CDW orders of these compounds to the saddle bands in the vicinity of their chemical potential. Even though such flat bands have been observed in these materials, they don't seem to appear at relevant energies \cite{OLSON_TaS2_PRB, OLSON_TaSe2_PRL, OLSON_TaSe2_PRB, BORISENKO_NbSe2_PRL, KYLE_NbSe2_PRB_OLD, UC_TaS2}. These theoretical approaches are essentially in the weak-coupling limit. Some of the intriguing experimental observations, for example the pseudogap behavior (that is persistence of non-zero CDW energy gap at temperatures much higher than \Tcdw), however, are suggestive of  strongly coupled CDWs in these systems \cite{UC_NATCOM, UC_TaS2, BORISENKO_NbSe2_PRL}. In this connection, there are a few strong-coupling theoretical models  \cite{JVW_NATCOM, JVW_PRB, VARMA_STRONG_COUPLING, AT_TaSe2_PRL, AT_NbSe2_PRB, GORKOV_PRB}, in most of which strong el-ph coupling and its momentum-anisotropy are critical to the CDW instability. Lately, several experimental studies have also pointed towards  nontrivial momentum-dependence of the el-ph coupling being key to the CDW order of $2H$ TMDs \cite{PLUMMER_PNAS, FRANK_NbSe2, DEVERAUX_PNAS}.

In the light of the aforementioned experimental and theoretical research, a comprehensive study of the el-ph coupling ($\lambda$) in 2$H$ TMDs with an emphasis on momentum ($\textbf{k}$) anisotropy of $\lambda$, is highly desirable towards developing a microscopic model of their CDW orders. One way to procure knowledge of the $\textbf{k}$-dependence of el-ph coupling of a system is to utilize high resolution ARPES data. Typically, the interaction between electronic excitations and any collective mode such as a phonon or a magnon would modify the ${bare}$ (i.e., at the absence of interactions) electronic dispersion and the electronic lifetime in a material \cite{ASHCROFT_MERMIN_BOOK}. The information about the collective mode is encoded in these modifications, which can be directly measured by employing ARPES. Experimental techniques such as tunneling spectroscopy or heat capacity measurements can also be incorporated for investigating collective modes of a solid. However, these techniques inherently render information averaged over the entire Fermi surface of the system \cite{EL_PH_OTHER_THAN_ARPES}. This is where ARPES, with its unique capability of energy-momentum resolved measurements, can be  crucial for shedding light on the the $\textbf{k}$-dependence of the electron-mode coupling.

In some of the past ARPES publications, thorough examinations of the impact of collective modes on the electronic degrees of freedom have been conducted on \TaSe and \NbSe \cite{KYLE_NbSe2_PRB, KYLE_TaSe2_PRB, TONY_NbSe2_PRL, TONY_TaSe2_PRL}. The collective modes in \NbSe , as seen by ARPES, was inferred to be phonons. However,  the identity of the mode in \TaSe is yet to be resolved \cite{KYLE_TaSe2_PRB, TONY_TaSe2_PRL}. Moreover, the el-ph coupling of \NbSe was found to have appreciable $\textbf{k}$-dependence. Surprisingly, such investigations are absent in case of  \TaS, which is a prominent member of the $2H$ TMD family and an extremely versatile material by itself. In this context, unlike the cases of \NbSe and \TaSe, ARPES investigations on \TaS are rather limited \cite{FENG_NbSe2_PRL, FENG_TaS2_JPCS, OLSON_TaS2_PRB}, where the main objective has been the study of CDW energy gap.
 As to the specifics of its electronic orders,  \TaS hosts a canonical incommensurate CDW order with \qcdw$\sim (2/3-0.02)\,\Gamma$M \cite{BS4} and the \Tcdw $\sim$75K \cite{CDW_TaS2_1, CDW_TaS2_2}. Intertwining of superconductivity and CDW at temperatures lower than 0.8K \cite{SC_TaS2_1, SC_TaS2_2, SC_TaS2_3}, combined with the influence of strong correlations among Ta $d$ electrons, leads to exceptionally rich phase diagrams of \TaS, which have drawn enormous scientific attention of late \cite{CAVA_Cu_TaS2, Cu_TaS2_2,  Fe_TaS2, Ni_TaS2, Na_TaS2_1, Na_TaS2_2, INTERCALATION_TaS2_1, INTERCALATION_TaS2_2,PRESSURE_TaS2, EXFOLIATION_TaS2}.

In this article, we have studied the possible interplay of collective modes and electronic excitations in \TaS using ARPES measurements. Such a study is anticipated to be vital not only to unveiling the mechanism of CDW order in \TaS , but also to deciphering the universal traits of incommensurate CDW systems in $2H$ TMDs, if there  are any.

\section{Experimental Details}
We have conducted ARPES measurements using the 21.2 eV Helium-I line of a discharge lamp combined with a Scienta R3000 analyzer at the University of Virginia, as well as 75 eV and 22 eV synchrotron light equipped with a Scienta R4000 analyzer at the PGM beamline of the Synchrotron Radiation Center, WI. The angular resolution is $\sim$ 0.3 degree, and the energy resolution is $\sim$ 8--15 meV. For temperature ($T$)-dependent studies, data were collected in a cyclic way to ensure that there were no aging effects in the spectra. All experiments were performed in ultra high vacuum (better than $5\times10^{-11}$Torr, both in the helium lamp system and in the beamline). The single crystal samples were cleaved {\it in situ} to expose their fresh surfaces for ARPES measurements. Samples were cooled using a closed cycle He refrigerator and the sample temperatures were measured using a silicon diode sensor mounted close to the sample holder. During each measurement the chemical potential  of the system was determined by fitting the ARPES spectrum of a polycrystalline gold reference sample, being in electrical contact with the sample under study, by Fermi Dirac distribution function. Single crystals of \TaS were grown using the standard iodine vapor transport method.

\begin{figure}[h]
\includegraphics[width=3.52in]{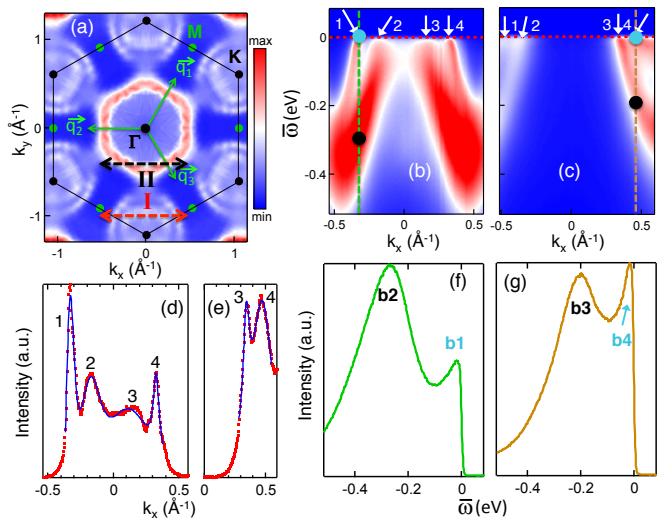}
\caption{(a) Fermi surface (FS) intensity map of a \TaS sample. Green arrows represent the three primary CDW vectors $\mathbf{q_1}$, $\mathbf{q_2}$ and $\mathbf{q_3}$. The method for constructing the FS intensity map is elaborated in the text.  EMIM along (b) momentum line I and (c) momentum line II.  The momentum lines I and II are marked by the red and black double-sided arrows in (a). Four Fermi crossings (FCs) in each of (b) and (c) are numbered from 1 to 4 and  they are also indicated using white arrows. The red dashed lines in (b) and (c) correspond to the chemical potential.
(d) MDC (shown by red markers) together with Voigt fit (shown by blue line) at \wbar $\sim$ 0 meV along momentum line I. (e) Same as (d) but along momentum line II. We have focussed on rightmost two peaks (associated with FC 3 and FC 4) in (e) since the intensities of the other two peaks (associated to FC 1 and  FC 2) are very weak.  (f) EDC at the momentum location shown by the green dashed line in (b). (g) Same as (f), but at the momentum location shown by the orange dashed line in (c). Double-peak structure of both EDCs signify bilayer-split bands in (b) and (c).}
\end{figure}
  
\section{Results}

\subsection{Low-energy electronic structure of \TaS}

In this section, we inspect various aspects of low energy electronic excitations in \TaS.  We begin by looking at Fig. 1(a), which depicts  the Fermi surface (FS) intensity map of \TaS in the normal state. The FS intensity map corresponds to the ARPES data $\text{I}(k_x, k_y, \overline{\omega})$ at $\overline{\omega}=0$ as a function of the in-plane momentum components ${k_{x}}$ and ${k_{y}}$, where $\overline{\omega}$ is the electronic energy referenced to the chemical potential. The necessary ARPES data were collected at $T\sim 90$K$>$\Tcdw($\sim 75$K) with photon energy $h\nu=$75eV.
As predicted by the first-principles calculations, the FS of \TaS consists of double-walled FS sheets around the $\Gamma$ point and the K points \cite{BS1,BS2, BS3, BS4}. This double-walled  nature of the FS sheets is due to the presence of two formula units per unit cell. 
The method for constructing the FS intensity map in Fig. 1(a) is as follows: starting from the raw ARPES data, we first subtract the constant signal at $\overline{\omega}$$>$0 (due to second order light) and then, we normalize each ARPES spectrum by the area enclosed by it and the energy axis between measured values of $\overline{\omega}$. The raw data covered more than one half of the surface BZ. For better visualization, the entire FS intensity map was constructed by reflections, using interpolations to uniform grids.

 The bilayer-split energy bands, which give rise to these double-walled FS sheets around $\Gamma$ and K points, can be better visualized from the ARPES energy-momentum intensity maps (EMIM's) in Figs. 1(b) (recorded along momentum line I) and 1(c) (recorded along momentum line II). Two distinct energy bands, separated in energy and momentum, are clearly visible on each side of the symmetry point of the EMIM in both cases. Note that these EMIMs are based on raw ARPES data collected using $h\nu=22$eV.
 It is worth mentioning that the FS as well as the EMIMs, similar to those seen here, have also been reported for \NbSe and \TaSe \cite{OLSON_TaS2_PRB, OLSON_TaSe2_PRL, BORISENKO_NbSe2_PRL, KYLE_NbSe2_PRB, KYLE_TaSe2_PRB}. 
 
One can also look into these bilayer-split bands using one-dimensional representations of ARPES data. For example, ARPES data as a function of one of the components of $\textbf {k}$ (either $k_x$ or $k_y$) for a given value of $\overline{\omega}$ is known as a momentum distribution curve (MDC). On the other hand, ARPES data as a function of $\overline{\omega}$ for a specific value of $\textbf {k}$ is known as an energy distribution curve (EDC). We display a representative MDC in Fig. 1(d) along the momentum line I at $\overline{\omega}\sim$ 0 meV. A glance at Fig. 1(b) shows that the momentum line I crosses through four bands at their individual Fermi crossings (FCs).  
 
 Therefore, the MDC in Fig. 1(d) exhibits four peaks. Similar data along the momentum line II has been presented in Fig. 1(e). For visual clarity, we have here focussed on only one pair of peaks , since the relative intensity of the other pair is extremely weak. 
 
 Additionally, two representative EDCs are presented in Figs. 1(f) and 1(g).The EDC in Fig. 1(f) is at the momentum location, marked by the green dashed line in Fig. 1(b). This line, which is at constant momentum, intersects the bilayer-split bands at two different values of \wbar. Hence, this EDC shows two peaks, namely b1 and b2. Similarly, the two peaks b3 and b4 of the EDC in Fig. 1(g) corroborate bilayer-split energy bands in Fig. 1(c) as well.

  Now we turn to the possible interconnection between  \qcdw and FS nesting vectors. Fig. 1(a) establishes that although the FS of \TaS has a number of nearly parallel regions, their separations do not agree with the magnitude of \qcdw. For instance, both FS sheets around the $\Gamma$ point are too large in size for being self-nested by any of the three primary CDW wave vectors $\mathbf{q_1}$, $\mathbf{q_2}$ and $\mathbf{q_3}$ (shown by green arrows in Fig. 1(a)). A simple FS nesting alone is therefore not likely to be the driver of the CDW instability in \TaS. Similar conclusion has also been reached in case of \NbSe \cite{BORISENKO_NbSe2_PRL}.

 \subsection{Multiple $kinks$ in band dispersion} 

The details of the interaction between a collective mode and the electronic excitations of a system is encapsulated in its single-particle self-energy $\Sigma(\textbf{k}, \overline{\omega})$, which is a complex-valued function of momentum $\textbf{k}$ and energy \wbar. The real part ${\Sigma}^{'}(\textbf{k},\overline{\omega})$ and the imaginary part ${\Sigma}^{''}(\textbf{k},\overline{\omega})$ manifest the renormalizations of the bare electronic dispersion and life-time respectively, due to many-body interactions \cite{MAHAN_BOOK}.

Both $\Sigma^{'}$  and $\Sigma^{''}$ may, in principal, be evaluated directly from ARPES spectra as ARPES intensity $\text{I}(\textbf{k}, \overline{\omega})$ can approximately be expressed as follows:  $\text{I}(\textbf{k}, \overline{\omega})=M(\textbf{k})\text{A}(\textbf{k},\overline{\omega})f(\overline{\omega})$, where (i) $f(\overline{\omega})$ is the Fermi Dirac distribution function, (ii) $M(\textbf{k})$ is the dipole matrix element, (iii) the spectral function $\displaystyle \text{A}(\textbf{k}, \overline{\omega})=\frac{\Sigma^{''} (\textbf{k}, \overline{\omega})}{(\overline{\omega}-\epsilon_\textbf{k}-\Sigma^{'} (\textbf{k},\overline{\omega}))^2+{\Sigma^{''}(\textbf{k}, \overline{\omega})}^2}$ and (iv) $\epsilon_\textbf{k}$ is the bare electronic dispersion \cite{HUFNER_BOOK, JC_REVIEW, ZX_REVIEW, JHONSON_REVIEW}. For the most generic case, obtaining $\Sigma^{'}$ and $\Sigma^{''}$ from ARPES data is a challenging task since  $\Sigma^{'}$ and $\Sigma^{''}$ are functions of both $\textbf{k}$ and $\overline{\omega}$. In such a case, one may have to assume ${a}$ ${ priori}$ some representation of the relevant many-body interaction, which often itself is a difficult job, for making progress in data analysis. However, the data analysis gets substantially simplified when $\Sigma$ and $M$ are independent of $\textbf{k}$ or even have weak $\textbf{k}$ dependence.

\begin{figure}[h]
\includegraphics[width=3.52in]{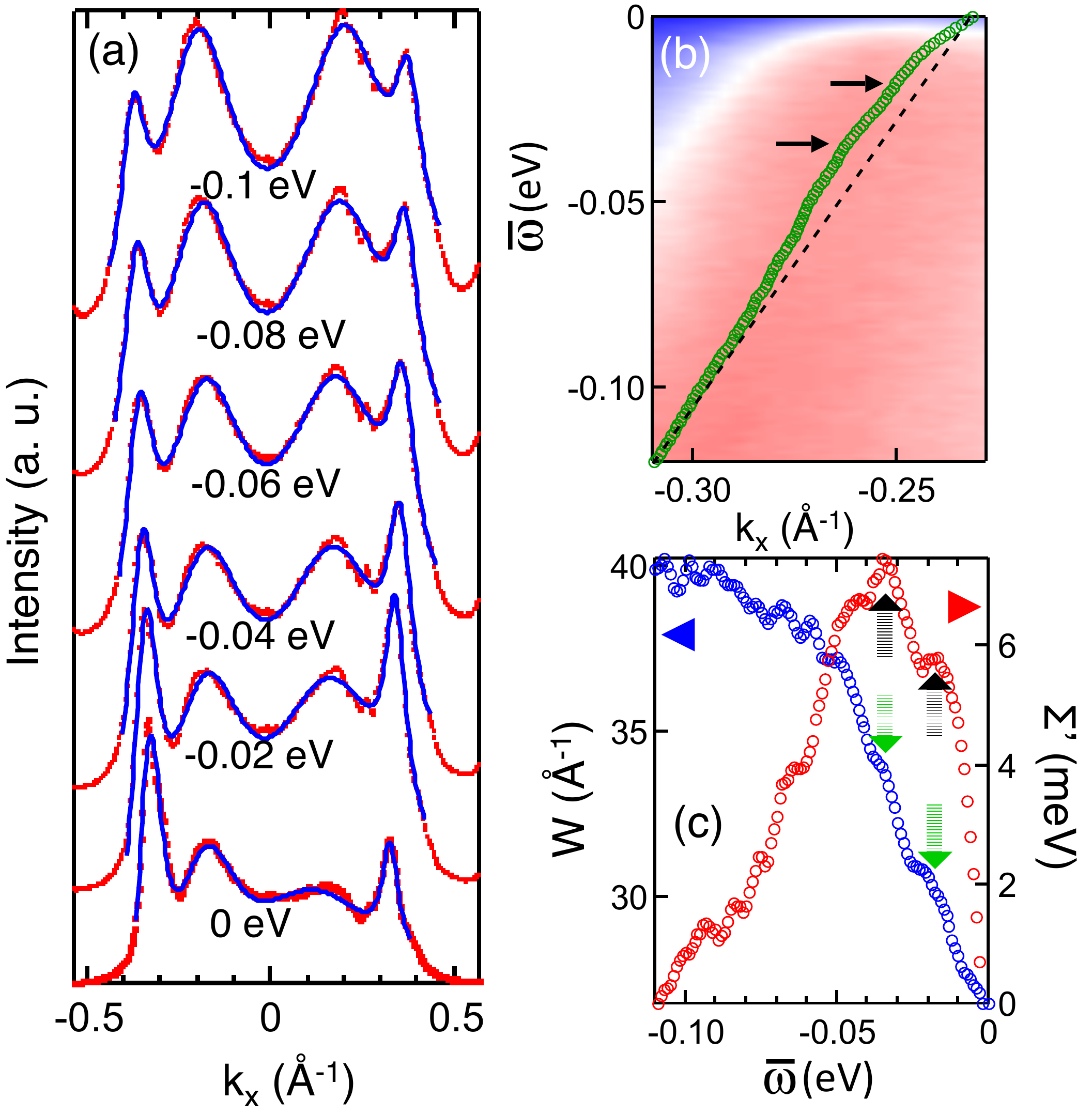}
\caption{ (a) MDCs (red markers) together with fits (blue lines) along the momentum line I (Fig. 1(a)) for a series of \wbar's. Note that the curves are offset for visual clarity. (b) Band dispersion (green circles) from MDC analysis is superimposed on the EMIM of Fig. 1(b) in the proximity of FC1. Approximated bare band dispersion is shown by black dashed line, while the kink locations are are marked with black arrows in (b). (c) Real part of the self-energy $\Sigma^{'}(\overline{\omega})$ (red circles, right axis), and the width of the MDC peaks $W(\overline{\omega})$(blue circles, left axis). Note that $\Sigma^{''}(\overline{\omega})$ is directly proportional to $W(\overline{\omega})$. Black dashed arrows indicate the peaks in $\Sigma^{'}(\overline{\omega})$, while the green ones point corresponding changes in slope of $W(\overline{\omega})$. As expected, \wbar locations of the black and the green dashed arrows approximately coincide.}
\end{figure}

In case of $\textbf k$-independent $\Sigma^{'}$, $\Sigma^{''}$ and $M$, MDCs at various values of $\overline{\omega}$ take simple Lorentzian line shape, at least in the vicinity of the Fermi momentum $k_F$ such that $\epsilon_\textbf{k}\sim v^0_F(|\textbf{k}|-k_F)$ with $v^0_F$ being the bare Fermi velocity. Lorentzian line shape of the MDC can be recognized from the above-mentioned form of $\text{A}(\textbf{k}, \overline{\omega})$. Consequently, the renormalized dispersion of an energy band can be determined by tracking the fitted peak locations of the corresponding MDCs as a function of \wbar.
 
Subsequently, $\Sigma^{'} (\overline{\omega})$ can be estimated by subtracting the bare dispersion from the renormalized dispersion \cite{HUFNER_BOOK, JC_REVIEW, ZX_REVIEW, JHONSON_REVIEW}. Moreover, $\Sigma^{''}(\overline{\omega})$ can be quantified from the fitted peak widths $W(\overline{\omega})$ of the MDCs. The relation between $\Sigma^{''}(\overline{\omega})$ and $W(\overline{\omega})$ is as follows: $\displaystyle W(\overline{\omega})=\frac{\Sigma^{''}(\overline{\omega})}{v_F^0}$ \cite{HUFNER_BOOK, JC_REVIEW, ZX_REVIEW, JHONSON_REVIEW}. One noticeable advantage of this analysis is that no particular model for the many-body interaction is necessary.

 MDCs at $\overline{\omega}\sim0$ along momentum lines I and II are displayed in Figs. 1(d) and 1(e) respectively. Notice that the Voigt function instead of the Lorentzian function was used for fitting these MDCs with the aim of approximately taking the momentum resolution into account. This in turn enables better estimates of the MDC peak widths. We now concentrate on the band dispersion using MDCs as a function of \wbar. In Fig. 2(a), we show the MDCs together with Voigt fits for a series of \wbar's along the momentum line I ( shown in Fig. 1(a)). In order to examine how the band dispersion, derived using the MDC analysis, correlate with the corresponding EMIM, we refer to Fig. 2(b). In Fig. 2(b), we superimpose the dispersion curve on the EMIM itself in the vicinity  of a FC, FC 1 (Fig. 1(c)) to be specific. The reason behind zooming into such a narrow \wbar-$k_x$ space of the EMIM in Fig. 1(b) is the fact that MDC analysis is strictly valid in the vicinity of the FC of an energy band. It can be readily observed that the MDC derived band dispersion agrees reasonably well with the intensity profile of the corresponding EMIM---at least in an energy window of 100 meV below the chemical potential.  A closer look at the dispersion curve in Fig. 2(b) further reveals that the band dispersion consists of multiple changes in slope, commonly referred as $kinks$.  Typically, the  presence of a kink in the electronic dispersion is interpreted as a fingerprint of the scattering of the electrons from some collective mode of the system. The obvious question is: can we  identify this mode?  To address this, we consult the published Raman scattering data \cite{RAMAN_SCATTERING_TaS2} as well as the theoretical calculation of the phonon spectra \cite{CALCUL_PHONON_TaS2_1, CALCUL_PHONON_TaS2_2}  of \TaS. The kink energies in the present case are $\sim$ 18 meV and $\sim$ 35 meV (shown by black solid arrows in Fig. 2(b)) and they agree well with the the optical phonon frequencies of \TaS. Therefore, it would be natural to conclude that interactions among electrons and optical phonons are responsible for the kinks in the dispersions. It is worth mentioning that similar multiple kinks of phononic origin have been reported in recent ARPES study of \NbSe \cite{KYLE_NbSe2_PRB}. Over the years,  kink features in electronic dispersion, may or may not be due to phonons, have also been observed in a wide array of solid state systems other than CDW materials, such as metallic sysrems \cite{METAL_KINK1, METAL_KINK2}, conventional superconductors \cite{ADAM_MgB2}, CMR manganites \cite{DAN_LSMO_PRL}, cuprate HTSCs \cite{JC_REVIEW, ZX_REVIEW, JHONSON_REVIEW}, and pnictide HTSCs \cite{HASAN_KINK_PNICTIDE}.

\subsection{${\Sigma}^{'}(\omega)$ and ${\Sigma}^{''}(\omega)$ from MDC analysis:} 
 The information about the bare band dispersion is necessary for estimating $\Sigma^{'}(\overline{\omega})$. We have approximated the bare band dispersion by fitting the high binding energy (\wbar$\sim$-100meV) part of the MDC derived dispersion with a straight line such that it crosses through $k_F$. 
 Similar procedure for estimating bare dispersion has been extensively employed in previous ARPES papers \cite{KYLE_NbSe2_PRB, JC_REVIEW, ZX_REVIEW, JHONSON_REVIEW, HASAN_KINK_PNICTIDE, DAN_LSMO_PRL, ADAM_MgB2}. The deviation of the experimentally observed deviation from the bare dispersion provides a measure for $\Sigma^{'}(\overline{\omega})$.

  Therefore, we quantify $\Sigma^{'}(\overline{\omega})$ by subtracting this approximated bare band dispersion from the measured one. Additionally, $\Sigma^{''}(\overline{\omega})$ can be obtained from $W(\overline{\omega})$. Both $\Sigma^{'}(\overline{\omega})$ and $W(\overline{\omega})$ are plotted in Fig. 2(c). It is apparent that $\Sigma^{'}(\overline{\omega})$ consists of peaks at \wbar's close to the kink energies found in Fig. 2(b). As expected from the requirements for causality, the peak energies of $\Sigma^{'}(\overline{\omega})$ approximately coincide with the \wbar locations of the relatively abrupt changes in $W(\overline{\omega})$ \cite{MAHAN_BOOK}.

\begin{figure}[h]
\includegraphics[width=3.52in]{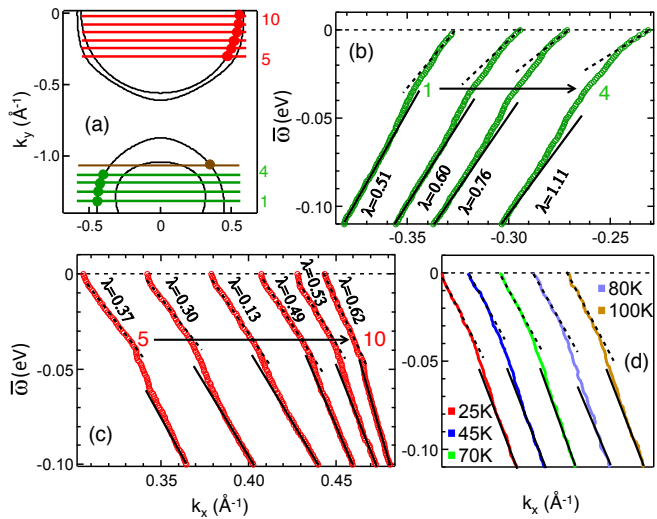}
\caption{ (a) Schematic plots of $\Gamma$- and K-centered FS sheets. Green and red solid lines show the momentum lines along which MDCs were constructed for extracting dispersions shown in (b) and (c). 
(b) MDC-derived band dispersions close to the FCs, which are marked by green solid dots and numbered from 1 to 4 in (a). (c) Same as (b) but close to the FCs, which are marked by red solid dots and numbered from 5 to 10 in (a).  Values of el-ph coupling parameter $\lambda$ at various FCs are also noted in (b) and (c). (d) MDC-derived band dispersions as a function of $T$ close to the FC, which is marked by the brown dot in (a). The brown solid line corresponds to the momentum line along which the pertinent MDCs were constructed. In each dispersion curve, the slope of the black dashed lines passing through the data points at \wbar$=0$ determine $v_F^*$, while the slope of the thick black line measures $v_F^0$.}
 \end{figure}
\subsection{$\textbf k$- and $T$-dependence of band renormalization:} 
In this section, we illustrate the $\textbf k$- and $T$-dependence of the band renormalization due to many-body interaction.  We will first examine $\textbf k$-dependence using Figs. 3(a), 3(b) and 3(c). As displayed in Fig. 3(a), we have focussed in the neighborhood of several FCs along the $\Gamma$-centered inner FS sheet and the K-centered outer FS sheet. We couldn't, however, obtain reliable band dispersions along the $\Gamma$-centered outer FS sheet and the K-centered inner FS sheet. Most probably, detailed photon energy sweep would be necessary to find the ideal photon energy for studying the quasiparticle dynamics of these FS sheets, which we plan to undertake in the future work. Fig. 3(b) summarizes the band dispersions along the K-centered outer FS sheet, while Fig. 3(c) along the $\Gamma$-centered inner FS sheet. Similar to the case in Fig. 2(b), each dispersion curve in Figs. 3(b) and 3(c) comprise of multiple kink structures. Within experimental error bars, the \wbar locations of these kinks reside within an energy window of 10meV$-$40meV.
 
This is in accord with the above-described proposition---the kinks in \TaS reflect the interactions among the electrons and various optical phonons.

One of the ways to quantify $\lambda$ is to employ the following relation: $\displaystyle \lambda=\frac{v_F^0}{v_F^*}-1$, where $v_F^0$ is the bare Fermi velocity, i.e., the slope of the approximated bare dispersion, and  $v_F^*$ is the renormalized Fermi velocity, i.e., the slope of the renormalized dispersion at the chemical potential \cite{JC_REVIEW, ZX_REVIEW, JHONSON_REVIEW}. Here the implicit assumption is that the entire renormalization effect is due to el-ph coupling. From Figs. 3(b) and 3(c), it can be deduced that $\lambda$ has a gentle $\textbf k$-dependence along both FS sheets under current consideration. The $\textbf k$-dependence, however, seems to be slightly more prominent along the K-centered FS sheet. 
We would like to point out that  $\textbf k$-dependent $\lambda$, somewhat more pronounced than \TaS, has also been reported for \NbSe \cite{TONY_NbSe2_PRL, KYLE_NbSe2_PRB}. Fig. 3(d) describes $T$-dependence of the renormalization effect. One can see that the kink structures in the dispersion remains intact even at temperatures higher than \Tcdw.  Furthermore, the kink energies are more or less $T$-independent. We would like to point that very similar results have been reported for \NbSe \cite{TONY_NbSe2_PRL}.These further support the el-ph interaction being the  driver of the kinks in the electronic dispersions of \TaS.

\section{Conclusions}
In summary, our systematic ARPES measurements on \TaS demonstrate dispersion renormalization, reflected by the deviations of the observed band dispersions from the bare ones, on K-centered outer and $\Gamma$-centered inner FS sheets. Comparison of the energy scales of the dispersion anomalies, i.e., the kinks, with the published data on phonon spectrum of \TaS reveals that interactions among electrons and optical phonons in the system give rise to these kinks. Similar to the case of \NbSe , the electron-phonon coupling parameter $\lambda$ of \TaS exhibits momentum anisotropy---$\lambda$ varies from 1.1 to 0.29, which is, however, weaker in comparison with the momentum-variation of $\lambda$ in \NbSe. In future works, it will be very interesting to explore whether this momentum-anisotropy of  $\lambda$  is manifested in terms of any momentum-dependence of CDW energy gap in \TaS.  Based on our study on \TaS and its correlation with the existing data on related 2H TMDs, 2H-NbSe2 for example, following two features seem to be universal to the incommensurate CDW order in TMDS---(i) lack of one-to-one correspondence between CDW wave vectors and the FS nesting vectors, and (ii) momentum-dependent el-ph coupling. 


~\\ \indent
{\bf Acknowledgments} \\
UC  acknowledge useful discussions with Patrick Vora and Jasper van Wezel. U.C. acknowledges supports from the National Science Foundation under Grant No. DMR-1454304 and from the Jefferson Trust at the University of Virginia. Work at Argonne National Laboratory (C.D.M., D.Y.C., M.G.K.) was supported by the U.S. Department of Energy, Office of Basic Energy Sciences, Division of Materials Science and Engineering.





\end{document}